\documentclass[a4paper,11pt]{article}
\usepackage{jheppub} 

\usepackage{soul}
\usepackage{tikz}
\usepackage{epsfig,latexsym,amsfonts,amsmath,amsthm,amssymb,amsbsy,multirow,slashed,wasysym,textcomp,subfigure,wrapfig,datetime,comment,mathtools,cancel,twistor,mathrsfs,empheq}
\usepackage[]{hyperref}
\usetikzlibrary{calc}

\newcommand{\csch}{\mathrm{csch}}

\usepackage[normalem]{ulem} 
\numberwithin{equation}{section}
\def\ee{\end{equation}}
\def\be{\begin{equation}}
\def\bea{\begin{eqnarray}}
\def\eea{\end{eqnarray}}
\newcommand{\beq}{\begin{eqnarray}}
\newcommand{\eqq}{\end{eqnarray}}
 \newcommand{\badat}{\begin{alignedat}}
 \newcommand{\eadat}{\end{alignedat}}

\newcommand{\eal}[1]{\be \begin{aligned} #1 \end{aligned}\end{equation}} 

\newcommand{\eqn}[1]{\be #1 \end{equation}} 
\newcommand{\eqa}[1]{\bea  #1\end{eqnarray}}

\newcommand{\eps}{\epsilon}
\renewcommand{\del}{\partial}

\renewcommand{\l}{\lambda}

\usepackage{color}

\newcommand{\pink}[1]{\textcolor{\pink}{#1}}

\definecolor{dblue}{rgb}{0.2,0.50,0.80}

\usepackage{xspace}

\renewcommand{\D}{\Delta}

\newcommand{\G}{\Gamma}

\renewcommand{\scri}{\mathscr{I}}

\newcommand{\talpha}{\tilde{\alpha}}
\newcommand{\tbeta}{\tilde{\beta}}

\catcode`,\active

\catcode`\,12

\arxivnumber{2605.03155} 

\title{Flat Space Physics from AdS Actions}

\author[a,b]{Walker Melton}

\affiliation[a]{Center for the Fundamental Laws of Nature, Harvard University, Cambridge, MA 02138, USA}
\affiliation[b]{Society of Fellows, Harvard University, Cambridge, MA 02138, USA}

\emailAdd{wmelton@fas.harvard.edu}

\abstract{Flat spacetimes are foliated by hyperbolic slices that are geometrically three-dimensional de Sitter or anti-de Sitter spaces. As such, it is possible to construct flat space holographic dualities by applying the AdS/CFT bulk-to-boundary dictionary slice by slice. In this work, we reduce 4D actions for massless scalars in both Minkowski space and Klein space and massive scalars in Minkowski space to actions on these 3D dS and AdS slices. In both Minkowski and Klein space, the reduced theories have a continuous spectrum of fields originating from the reduction over the noncompact $x^2$ direction. These actions are linked by boundary terms arising from field configurations discontinuous across the lightcone. In the massless case, different asymptotic limits of the reduced field near the boundary of the unit hyperbolic slice replicate either lightcone or null infinity limits of the field; in the massive case, only one boundary mode of the reduced field has a simple geometric interpretation.}

\begin{document}
\maketitle
\flushbottom

\section{Introduction}
Holographic dualities provide an invaluable tool for studying quantum gravity in asymptotically anti-de Sitter (AdS) spacetimes. By relating quantum gravity in an asymptotically-AdS background to a lower dimensional CFT, holography has allowed us to  understand the fundamental degrees of freedom of quantum gravitational physics in AdS spacetimes and has given us concrete tools to study processes such as black hole evaporation where quantum gravitational effects are significant. 

However, it is difficult to apply the AdS/CFT correspondence to more physically realistic contexts because the extrapolate dictionaries depend sensitively on the large scale structure of anti-de Sitter spacetimes, relying either on taking limits where bulk operators approach the conformal boundary of AdS or on turning on non-trivial boundary conditions \cite{Banks:1998dd,Gubser:1998bc}. Because both dictionaries rely on the existence of a timelike conformal boundary, it is difficult to naively construct a similar dictionary relating gravity in asymptotically flat spacetimes, which have a null conformal boundary, or de Sitter spacetimes,  which have a spacelike conformal boundary, to a lower-dimensional boundary theory. 

While much progress has been made defining a holographic dictionary for flat spacetimes by combining general symmetry principles with  inspiration from AdS/CFT \cite{Pasterski:2016qvg, Pasterski:2017kqt, Mason:2023mti}, there are also several promising avenues for relating flat space holography with the AdS/CFT correspondence. Most obviously, flat Minkowski spacetime is the $\ell_{\mathrm{AdS}}\to\infty$ limit of AdS spacetime; in this limit, flat space scattering amplitudes can be extracted from perturbative holographic correlation functions \cite{Gary:2009ae, deGioia:2024yne, Marotta:2024sce, Alday:2024yyj}. From a boundary perspective, this corresponds to taking a Carrollian limit ($c \to 0$) of the dual theory, and this limit has been used to obtain Carrollian amplitudes from the flat space limit of the ABJM duality \cite{Lipstein:2025jfj}. 

Flat space holographic dualities may also be embedded in higher-dimensional AdS/CFT dualities. Double holography proposes that CFTs with a boundary are dual to gravitational systems with dynamical `end of the world' branes cutting off the bulk geometry. By tuning the tension of the EoW brane, one can describe a flat brane embedded in higher-dimensional AdS \cite{Hao:2025ocu}. One can also obtain flat space dualities by relating theories in a deformed twistor space (which is geometrically $S^3$ fibered over AdS$_3$) to lower-dimensional gravity through the Penrose transform \cite{Costello:2022jpg}.

Flat space holographic dualities may also be built up from lower-dimensional AdS dualities. In four dimensions, Minkowski space is foliated by Euclidean AdS$_3$ and Lorentzian dS$_3$ slices, and one can generate holographic correlators by applying the appropriate dictionary slice by slice. This procedure has been used to construct correlation functions living on a pair of 2-spheres \cite{deBoer:2003vf, Solodukhin:2004gs}, study the extended BMS group \cite{Ball:2019atb}, and define celestial correlation functions from radial Mellin transforms of time-ordered correlation functions \cite{Sleight:2023ojm, Iacobacci:2024nhw}, and extract holographic correlators from 4D on-shell actions after an appropriate holographic renormalization \cite{Hao:2023wln}. Similar foliations have also been used to study MHV gluon amplitudes in nontrivial scalar backgrounds \cite{Casali:2022fro},  de-singularize celestial amplitudes into celestial leaf amplitudes \cite{Melton:2023bjw} and construct simple 2D CFTs whose correlation functions replicate these leaf amplitudes in the classical large-$N$ limit \cite{Melton:2024akx}.

In this paper, we continue this program by explicitly relating off-shell flat space actions for massless and massive particles to  off-shell actions living on the hyperbolic slices of Minkowski or Klein space. The resulting actions have a noncompact set of fields with masses chosen so that the dual conformal weights always lie on the principal series. Along with an off-diagonal free action arising from the action in the interior and exterior of the lightcone, the reduced action contains a boundary term generated by evaluating the bulk action on field configurations discontinuous across the lightcone of the origin. This boundary term is necessary to correctly describe propagation across the lightcone and recover 4D correlation functions from the 3D reduced action. 

The structure of the reduced action has several important consequences. First, quantizing the reduced theory in the appropriate time direction on the 3D slice may be physically inequivalent to more familiar quantization schemes; in Klein space, the time variable on each AdS$_3/\mathbb{Z}$ slice is an angular coordinate in 4D, so the resulting quantization may define a new quantization that may be more amenable to a holographic description while still reproducing the massless $\mathcal{S}$-matrix \cite{Melton:2024pre, Chen:2025acl}. The lightcone terms in the reduced action allow us to replicate bulk-to-bulk exchange, which will be essential for extending leaf dualities beyond the MHV sector \cite{Melton:2023bjw, Melton:2024akx}. Understanding the structure of the boundary terms will also be crucial for understanding how translation invariance emerges from the combining the duals of each spacetime region, which are not by themselves invariant under spacetime translations \cite{Melton:2024pre}.

Additionally, we study how the reduced field variables behave under the boundary limits on each hyperbolic slice, such as the extrapolate limits considered in \cite{deBoer:2003vf} and \cite{Sleight:2023ojm}. For massless fields, the extrapolate limits of the reduced 3D field are equivalent to taking a limit where the 4D field approaches the lightcone or null infinity, recasting the normalizable and non-normalizable boundary conditions on the reduced field as values of the 4D field on the lightcone and null infinity. This replicates the natural holographic dictionary in flat space \cite{Ruzziconi:2026bix}.  For massive scalars, however, one linear combination of the possible extrapolate limits encodes the value of the field on the lightcone. The other limit seemingly has no simple geometric interpretation. Understanding what the 3D extrapolate limit of the reduced field computes will hopefully lead to a deeper understanding of the role of massive particles in flat space holography. 

This paper is organized as follows. After reviewing the hyperbolic foliations of Minkowski and Klein space in Section \ref{sec:bkgd}, we study the reduction of a massless scalar field in Section \ref{sec:massless}. We understand the extrapolate limits of the reduced field, compute the reduced action including the appropriate boundary terms in both Minkowski and Klein space, describe how non-derivative interactions can be included, and show that the reduced action correctly generates the 4D equations of motion and correlation functions. Section \ref{sec:massive} computes the extrapolate limits and reduced action for a free massive scalar in Minkowski space. 

\section{Background}\label{sec:bkgd}
In this section, we review the hyperbolic foliation of flat Minkowski and Klein spacetimes.
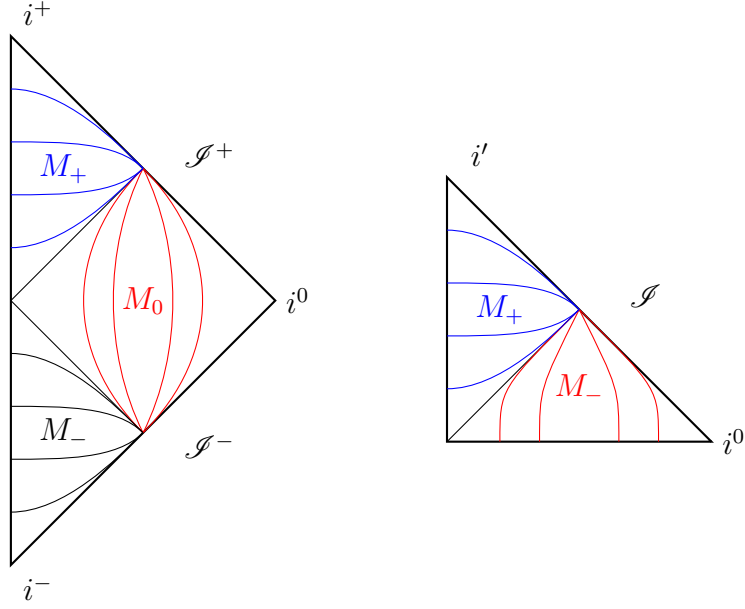
\begin{figure}[h]
\begin{center}
\begin{tikzpicture}[scale=3.5]

  \draw[thick] (0,-1) -- (0,1) -- (1,0) -- cycle;


  \node[above left] at (0.2,1) {$i^+$};
  \node[below left] at (0.2,-1) {$i^-$};
  \node[right] at (1,0) {$i^0$};

  \node[] at (0.75,0.55) {$\mathscr{I}^+$};
  \node[] at (0.75,-0.55) {$\mathscr{I}^-$};

    \draw (0,0) -- (0.5,0.5);
    \draw (0,0) -- (0.5,-0.5);
  \draw[red] (0.5,0.5) .. controls (.2,.2) and (.2,-.2)  .. (0.5,-0.5);
\draw[red] (0.5,0.5) .. controls (.35,.2) and (.35,-.2) .. (0.5,-0.5);
\draw[red] (0.5,0.5) .. controls (.65,.2) and (.65,-.2)  .. (0.5,-0.5);
\draw[red] (0.5,0.5) .. controls (.8,.2) and (.8,-.2)  .. (0.5,-0.5);
\draw[blue] (0.5,0.5) .. controls (.4,.4) and (.2,.4) .. (0,.4);
\draw[blue] (0.5,0.5) .. controls (.4,.4) and (.2,.2) .. (0,.2);
\draw[blue] (0.5,0.5) .. controls (.4,.6) and (.2,.6) .. (0,.6);
\draw[blue] (0.5,0.5) .. controls (.4,.6) and (.2,.8) .. (0,.8);
\draw[] (0.5,-0.5) .. controls (.4,-.4) and (.2,-.4) .. (0,-.4);
\draw[] (0.5,-0.5) .. controls (.4,-.4) and (.2,-.2) .. (0,-.2);
\draw[] (0.5,-0.5) .. controls (.4,-.6) and (.2,-.6) .. (0,-.6);
\draw[] (0.5,-0.5) .. controls (.4,-.6) and (.2,-.8) .. (0,-.8);
\node[blue] at (0.2,0.5) {$M_+$};
\node[] at (0.2,-0.5) {$M_-$};
\node[red ] at (0.5, 0.0) {$M_0$};
\end{tikzpicture}
\hspace{.5in}
\begin{tikzpicture}[scale=3.5,]

  \draw[thick] (0,0) -- (0,1) -- (1,0) -- cycle;


  \node[above left] at (0.2,1) {$i'$};
    \node at (0,-.6) {};
  \node[right] at (1,0) {$i^0$};

  \node[] at (0.75,0.55) {$\mathscr{I}$};

    \draw (0,0) -- (0.5,0.5);
  \draw[red] (0.5,0.5) .. controls (.2,.2) and (.2,.2)  .. (0.2,0);
\draw[red] (0.5,0.5) .. controls (.35,.2) and (.35,.2) .. (0.35,0);
\draw[red] (0.5,0.5) .. controls (.65,.2) and (.65,.2)  .. (0.65,0);
\draw[red] (0.5,0.5) .. controls (.8,.2) and (.8,.2)  .. (0.8,0);
\draw[blue] (0.5,0.5) .. controls (.4,.4) and (.2,.4) .. (0,.4);
\draw[blue] (0.5,0.5) .. controls (.4,.4) and (.2,.2) .. (0,.2);
\draw[blue] (0.5,0.5) .. controls (.4,.6) and (.2,.6) .. (0,.6);
\draw[blue] (0.5,0.5) .. controls (.4,.6) and (.2,.8) .. (0,.8);

\node[blue] at (0.2,0.5) {$M_+$};
\node[red ] at (0.5, 0.2) {$M_-$};
\end{tikzpicture}
\end{center}
\caption{A Penrose diagram of Minkowski space (left) and Klein space (right) depicting the hyperbolic foliation. In Minkowski space, $M_+$ is the interior of the future-pointing lightcone and is foliated by Euclidean AdS$_3$ slices, $M_0$ is the exterior of the lightcone and is foliated by Lorentzian dS$_3$ slices, and $M_-$ is the interior of the past-pointing lightcone and is foliated by Euclidean AdS$_3$ slices. In Klein space, both the timelike region ($M_+$) and the spacelike region ($M_-$) are foliated by Lorentzian AdS$_3/\mathbb{Z}$ slices. \label{fig:fig1}}
\end{figure}
\subsection{Minkowski Space}

We work in four-dimensional Minkowski space $\mathbb{M}^{1,3}$ with metric
\begin{equation}
    ds^2 = -(dx^0)^2 + (dx^1)^2 + (dx^2)^2 + (dx^3)^2 .
\end{equation}
It is convenient to foliate $\mathbb{M}^{1,3}$ by three-dimensional hyperbolic slices in both the timelike and spacelike regions separated by the lightcone $x^2 \equiv \eta_{\mu\nu}x^\mu x^\nu=0$. We will use Greek indices to represent full Minkowski space indices, while Latin indices will describe indices on the constant $x^2$ slice and will be raised and lowered with the induced metric on the slice.  The geometry of Minkowski space is depicted on the left side of Figure \ref{fig:fig1}. 

\paragraph{Timelike region ($x^2<0$).}
Inside the future lightcone ($x^0>0$, $x^2<0$) we introduce coordinates
\begin{equation}
    x^\mu = \tau\,\hat{x}_+^\mu,
    \qquad
    \hat{x}_+^\mu=(\cosh\zeta,\sinh\zeta\,\hat n),
    \qquad
    \hat n^2=1,
\end{equation}
with $\tau\in(0,\infty)$ and $\zeta\in[0,\infty)$. The metric becomes
\begin{equation}
    ds^2 = -d\tau^2+\tau^2\,ds_{H_3}^2,
    \qquad
    ds_{H_3}^2=d\zeta^2+\sinh^2\zeta\,d\Omega_2^2,
\end{equation}
where $ds_{H_3}^2$ is the metric on unit hyperbolic space H$_3$ (Euclidean AdS$_3$).

Inside the past lightcone ($x^0<0$, $x^2<0$) we similarly write, with $\tau \in (-\infty,0)$ so that $\tau$ is a forward pointing coordinate,
\begin{equation}
    x^\mu=\tau\,\hat{x}_-^\mu,
    \qquad
    \hat{x}_-^\mu=(\cosh\zeta,-\sinh\zeta\,\hat n),
\end{equation}
for which the metric again takes the form
\begin{equation}
    ds^2=-d\tau^2+\tau^2\,ds_{H_3}^2 .
\end{equation}
For convenience, we define $\tilde{\tau} = -\tau$ in the interior of the past-pointing lightcone. 
\paragraph{Spacelike region ($x^2>0$).}
Outside the lightcone we use coordinates
\begin{equation}
    x^\mu=\rho\,\hat{x}_0^\mu,
    \qquad
    \hat{x}_0^\mu=(\sinh\eta,\cosh\eta\,\hat n),
\end{equation}
with $\rho\in(0,\infty)$ and $\eta\in(-\infty,\infty)$. In these coordinates,
\begin{equation}
    ds^2 = d\rho^2 + \rho^2\,ds_{dS_3}^2,
    \qquad
    ds_{dS_3}^2 = -d\eta^2 + \cosh^2\eta\,d\Omega_2^2,
\end{equation}
so constant $\rho$ slices are Lorentzian dS$_3$ of unit radius.

\paragraph{Lightcone.}
The future and past lightcones can be parametrized as
\begin{equation}
    x^\mu=r\,q_\pm^\mu,
    \qquad
    q_\pm^\mu=(\pm 1,\hat n),
    \qquad
    r\ge 0.
\end{equation}

\subsection{Approaching the Lightcone}

To approach the future lightcone from inside the future timelike region, take
\begin{equation}
    \tau\to 0,\qquad \zeta\to\infty,\qquad \tau e^{\zeta}=2r\ \ \text{fixed},
\end{equation}
so that
\begin{equation}
    \lim_{\substack{\tau\to 0\\ \tau e^{\zeta}=2r}}
    \tau\,\hat{x}_+^\mu
    =
    r(1,\hat n)
    =
    r\,q_+^\mu .
\end{equation}
From the spacelike region, the future lightcone is reached by taking
\begin{equation}
    \rho\to 0,\qquad \eta\to\infty,\qquad \rho e^{\eta}=2r\ \ \text{fixed},
\end{equation}
which yields
\begin{equation}
    \lim_{\substack{\rho\to 0\\ \rho e^{\eta}=2r}}
    \rho\,\hat{x}_0^\mu
    =
    r(1,\hat n)
    =
   r\,q_+^\mu .
\end{equation}
Conversely, the past lightcone is reached from the spacelike region by taking
\begin{equation}
    \rho\to 0,\qquad \eta\to-\infty,\qquad \rho e^{-\eta}=2r\ \ \text{fixed},
\end{equation}
so that
\begin{equation}
    \lim_{\substack{\rho\to 0\\ \rho e^{-\eta}=2r}}
    \rho\,\hat{x}_0^\mu
    =
    r(-1,\hat n)
    =
    r\,q_-^\mu .
\end{equation}
Finally, approaching the past lightcone from inside the past timelike region is achieved by
\begin{equation}
    \tau\to 0,\qquad \zeta\to\infty,\qquad -\tau e^{\zeta}=2r\ \ \text{fixed},
\end{equation}
which gives
\begin{equation}
    \lim_{\substack{\tau\to 0\\ -\tau e^{\zeta}=2r}}
    \tau\,\hat{x}_-^\mu
    =
   r(-1,\hat n)
    =
   r\,q_-^\mu .
\end{equation}

\subsection{Approaching Null Infinity}

Introduce Bondi coordinates
\begin{equation}
    x^0 = u + r,
    \qquad
    \vec{x}=r\,\hat n,
\end{equation}
so that future null infinity $\mathscr{I}^+$ is reached by taking $r\to\infty$ at fixed retarded time $u=x^0-r$, while past null infinity $\mathscr{I}^-$ is reached by taking $r\to\infty$ at fixed advanced time $v=x^0+r$. Massless scalar fields fall off as $1/r$ at null infinity, giving 
\begin{equation}
    \begin{split}
        \phi(u,r,\hat{n}) & \sim \frac{\phi^{(1)}(u,\hat{n})}{r} + O(1/r^2)\ \mathrm{near}\ \mathscr{I}^+ \\
        \phi(v,r,\hat{n}) & \sim\frac{\phi^{(1)}(v,\hat{n})}{r} + O(1/r^2)\ \mathrm{near}\ \mathscr{I}^-.
    \end{split}
\end{equation}

Inside the future lightcone, one finds
\begin{equation}
    u = \tau(\cosh\zeta-\sinh\zeta)=\tau e^{-\zeta},
    \qquad
    r=\tau\sinh\zeta .
\end{equation}
Thus the portion of $\mathscr{I}^+$ with $u>0$ is reached by taking $\tau\to\infty$ with $u=\tau e^{-\zeta}$ held fixed.

The complementary portion of $\mathscr{I}^+$ with $u<0$ is reached from the spacelike region. Using
\begin{equation}
    u = x^0-r=\rho(\sinh\eta-\cosh\eta)=-\rho e^{-\eta},
\end{equation}
we reach $u<0$ by taking $\rho\to\infty$ with $u=-\rho e^{-\eta}$ held fixed.

Similarly, past null infinity is obtained at $r\to\infty$ with $v=x^0+r$ fixed. From the spacelike region,
\begin{equation}
    v = x^0+r=\rho(\sinh\eta+\cosh\eta)=\rho e^{\eta},
\end{equation}
so the half of $\mathscr{I}^-$ with $v>0$ is reached by $\rho\to\infty$ at fixed $v=\rho e^{\eta}$.
The remaining half with $v<0$ is reached from inside the past lightcone, where
\begin{equation}
    v=x^0+r=-\tau(-\cosh\zeta+\sinh\zeta)=\tau e^{-\zeta},
\end{equation}
and hence corresponds to $\tau\to-\infty$ with $v=\tau e^{-\zeta}$ fixed.

\subsection{Klein Space}

(2,2) signature Klein space has metric
\begin{equation}
    ds^2 = -(dx^0)^2 - (dx^1)^2 + (dx^2)^2+(dx^3)^2.
\end{equation}
In the region where $x^2 < 0$, we choose coordinates with 
\begin{equation}
    x^\mu = \tau \hat{x}_+^\mu,\ \hat{x}_+ = (\cosh\rho\cos\psi,\cosh\rho\sin\psi,\sinh\rho\cos\phi,\sinh\rho\sin\phi).
\end{equation}
In the spacelike region we have
\begin{equation}
    x^\mu = \tau\hat{x}_-^\mu,\ \hat{x}_- = (\sinh\rho\cos\phi,\sinh\rho\sin\phi,\cosh\rho\cos\psi,\cosh\rho\sin\psi).
\end{equation}
These coordinates are chosen so that the timelike and spacelike regions are related by exchanging $(x^0,x^1)$ with $(x^2,x^3)$.

The lightcone, which now has a single connected component, is reached from the timelike region by sending
\begin{equation}\label{eq:kslclt}
   \lim_{\substack{\tau\to 0\\ e^\rho\tau = 2r}} \tau\hat{x}_+^\mu = r(\cos\psi,\sin\psi,\cos\phi,\sin\phi).
\end{equation}
It is reached from the spacelike region by sending
\begin{equation}\label{eq:kslcls}
    \lim_{\substack{\tau\to 0\\ e^\rho\tau = 2r}} \tau\hat{x}_-^\mu = r(\cos\phi,\sin\phi,\cos\psi,\sin\psi).
\end{equation}
Note that the timelike and spacelike angles are reversed. 

Conversely, null infinity is reached by sending $\rho$ to infinity and $\tau \to \infty$ while fixing $\tau e^{-\rho}$. In the interior of the timelike wedge, we reach the region of $\mathscr{I}$ with $u > 0$ by sending $\rho \to \infty$ while fixing $u = \tau(\cosh\rho - \sinh\rho) = \tau e^{-\rho}$. The region of $\mathscr{I}$ with $u < 0$ is found from the spacelike wedge by sending $\rho \to \infty$ while fixing $u = -\tau e^{-\rho}$. 
\section{Massless Scalars}\label{sec:massless}
First, we consider the reduction of the free massless scalar, with action 
\begin{equation}
    S = -\frac{1}{2}\int d^4x \partial^\mu\phi\partial_\mu\phi.
\end{equation}
\subsection{Minkowski Space}
In this section we compute the reduced action for a massless scalar in Minkowski space. The resulting action will have the form of a pair of actions on H$_3$ and one on dS$_3$ coupled by boundary terms. 
\subsubsection{Reduction of Field Operators}

To reduce the action to a sum of actions on unit hyperbolic slices, we relate the field to an integral of 3D fields
\begin{equation}
    \phi = \begin{cases}
        \int [d\Delta]T_\l(\tau)\phi^+_\l(\zeta,\hat{n}) & x \in M_+ \\
        \int [d\Delta] R_\l(\rho)\phi^0_\l(\eta,\hat{n}) & x \in M_0 \\
        \int [d\Delta] T_\l(-\tau)\phi^-_\l(\zeta,\hat{n}) & x \in M_-
    \end{cases}.
\end{equation}
Here $\D = 1+i\l$ and $\int [d\D] = \int_{-\infty}^\infty d\l/2\pi$.                       In order for the action to reduce to a simple action on each hyperbolic slice, we want to choose the radial profiles $T_\l, R_\l$ so that the massless wave equation is equivalent to the massive wave equations on the hyperbolic slice: 
\begin{equation}
    \square \phi = 0 \Leftrightarrow (\square_{H_3} + (1+\lambda^2))\phi^\pm_\l = 0,\  (\square_{dS_3} - (1+\lambda^2))\phi^0_\l = 0.
\end{equation}
These equations imply that the radial profile functions obey
\begin{equation}
    \begin{split}
        \tau^2 T_\l''(\tau) + 3\tau T'_\l(\tau) + (1+\l^2)T_\l &= 0 \\
        \rho^2 R_\l''(\rho) + 3\rho R'_\l(\rho) + (1+\l^2)R_\l &= 0.
    \end{split}
\end{equation}
A convenient choice is $T_\l = \tau^{-1-i\l}, R_\l = \rho^{-1-i\l}$, so that we reduce the field by setting 
\begin{equation}
    \phi(x) = \begin{cases}
        \int[d\D] \tau^{-1-i\lambda}\phi^+_\l(\zeta,\hat{n}) & x \in M_+ \\
        \int[d\D] \rho^{-1-i\l}\phi^0_\l(\eta,\hat{n}) & x \in M_0 \\
        \int [d\D] (-\tau)^{-1-i\l}\phi^-_\l(\zeta,\hat{n}) & x \in M_- 
    \end{cases}.
\end{equation}
This can be inverted as 
\begin{equation}
    \begin{split}
        &\phi^+_\l = \int_0^\infty d\tau\tau^{i\lambda}\phi(\tau\hat{x}_+) \\
        &\phi^0_\l = \int_0^\infty d\rho\rho^{i\lambda}\phi(\rho\hat{x}_0) \\
        &\phi^-_\l = \int_0^\infty d\tilde{\tau}\tilde{\tau}^{i\l}\phi(-\tilde{\tau}\hat{x}_-).
    \end{split}
\end{equation}
\subsubsection{Extrapolate Limits of the Reduced Field}
In this section, we show that the 3D extrapolate limits of the reduced field correspond to lightcone or null infinity limits of the 4D field. 

First, consider the interior of the future-directed lightcone. In this region, one extrapolate limit becomes
\begin{equation}
    \begin{split}
        \lim_{\zeta\to\infty}e^{\D\zeta}\phi^+_\l(\zeta,\hat{n}) &= \lim_{\zeta\to\infty}\int_0^\infty \frac{d\tau}{\tau}(\tau e^{\zeta})^\D\phi(\tau\hat{x}_+) \\
        &= \lim_{\zeta\to\infty}\int_0^\infty\frac{dr}{r}(2r)^\D\phi(2re^{-\zeta}\hat{x}_+) \\
        &= \int_0^\infty\frac{dr}{r}(2r)^\D\phi(r q_+).
    \end{split}
\end{equation}
The integral over $\tau$ localizes to the region where $\tau \to 0$ as $\zeta\to \infty$, so that the extrapolate limit of the reduced field is a Mellin transform of the 4D field on the lightcone.

Similarly, the other extrapolate limit can be computed as
\begin{equation}
    \begin{split}
        \lim_{\zeta\to\infty}e^{(2-\D)\zeta}\phi^+_\l(\zeta,\hat{n}) &= \lim_{\zeta\to\infty} e^{2\zeta}\int_0^\infty \frac{d\tau}{\tau}(\tau e^{-\zeta})^{\Delta}\phi(\tau\hat{x}_+) \\
        &= 2\int_0^\infty du u^{\Delta-2}\phi^{(1)}(u,\hat{n})
    \end{split}
\end{equation}
so that taking the other possible extrapolate limit localizes to a Mellin transform of a field on null infinity. 

Inverting this and extending to the other regions of spacetime gives
\begin{equation}
    \begin{split}
        \phi^+_{L+}(r,\hat{n}) &= \lim_{\zeta\to\infty}\int [d\D](2r)^{-1-i\l}e^{(1+i\l)\zeta}\phi^+_\l(\zeta,\hat{n}) \\
        \phi^0_{L+}(r,\hat{n}) &= \lim_{\eta\to\infty} \int [d\Delta] (2r)^{-1-i\l}e^{(1+i\l)\eta}\phi^0_\l(\eta,\hat{n}) \\
        \phi^0_{L-}(r,\hat{n}) &= \lim_{\eta\to -\infty}\int [d\D](2r)^{-1-i\l}e^{-(1+i\l)\eta}\phi^0_\l(\eta,\hat{n}) \\
        \phi^-_{L-}(r,\hat{n}) &= \lim_{\zeta\to\infty}\int [d\D](2r)^{-1-i\l}e^{(1+i\l)\zeta}\phi^-_\l(\zeta,\hat{n}).
    \end{split}
\end{equation}

Similarly, we have that the values of the field on null infinity are described by 
\begin{equation}
    \begin{split}
    \phi^{(1)}(u > 0 ,\hat{n}) &=  \lim_{\zeta\to \infty}\frac{1}{2}\int [d\D]u^{-i\lambda}e^{(1-i\lambda)\zeta}\phi^+_\lambda(\zeta,\hat{n})\\
        \phi^{(1)}(u < 0, \hat{n}) &= \lim_{\eta\to\infty}\frac{1}{2}\int [d\D](-u)^{-i\l}e^{(1-i\l)\eta}\phi^0_\l(\eta,\hat{n}) \\
        \phi^{(1)}(v>0,\hat{n})  &= \lim_{\eta\to-\infty}\frac{1}{2}\int [d\D] v^{-i\l}e^{-(1-i\l)\eta}\phi^0_\l(\eta,\hat{n}) \\
        \phi^{(1)}(v < 0,\hat{n}) &= \lim_{\zeta\to\infty}\frac{1}{2}\int [d\D](-v)^{-i\l}e^{(1-i\l)\zeta}\phi^-_\l(\zeta,\hat{n}).
    \end{split}
\end{equation}

Hence, the values of the field on null infinity and the lightcone are controlled by the asymptotics of the reduced field near the boundary of the H$_3$ or dS$_3$ slice. At large $\zeta$, $\phi^+_\lambda$ can be expanded
\begin{equation}
    \phi^+_\lambda = e^{-\Delta\zeta}\alpha_\lambda^+(\hat{n}) + e^{(\Delta-2)\zeta}\beta_\lambda^+(\hat{n}).
\end{equation}
Let 
\begin{equation}
    \begin{split}
    A^+(\gamma,\hat{n}) &= \int[d\Delta] \gamma^{-\Delta}\alpha^+_\lambda(\hat{n}) \\
    B^+(\gamma,\hat{n}) &= \int [d\Delta]\gamma^{-\Delta}\beta^+_\lambda(\hat{n}).
    \end{split}
\end{equation}
Then, the lightcone limit of the field takes the form 
\begin{equation}\label{eq:lclimit}
    \begin{split}
        \phi^+_{L+}(r,\hat{n}) &= \lim_{\zeta\to\infty}\int [d\Delta](2r)^{-\Delta}[\alpha_\lambda^+(\hat{n}) + e^{2(\Delta-1)\zeta}\beta^+_\lambda(\hat{n})] \\
        &= A^+(2r,\hat{n}) + \lim_{\zeta\to \infty} e^{-2\zeta} B^+(e^{-2\zeta}2r,\hat{n}).
    \end{split}
\end{equation}
Conversely, near null infinity, we have that 
\begin{equation}
\begin{split}
    \phi^{(1)}(u>0,\hat{n}) &= \frac{u}{2}\int [d\Delta] u^{-\Delta}\lim_{\zeta\to \infty}e^{(2-\Delta)\zeta}\phi^+_\lambda(\zeta,\hat{n}) \\
    &= \frac{u}{2}\lim_{\zeta\to \infty}\int [d\Delta]u^{-\Delta}[\beta^+_\lambda + e^{2(1-\Delta)\zeta}\alpha^+_\lambda(\hat{n})] \\
    &= \frac{u}{2}B^+(u,\hat{n}) + \frac{u}{2}\lim_{\zeta\to \infty} e^{2\zeta}A^+(e^{2\zeta}u,\hat{n}).
    \end{split}
\end{equation}
There is an ambiguity in the formula because at $\Delta = 1$, the falloffs degenerate and we can shift the mode expansion. This allows us to set the limiting term in Equation \eqref{eq:lclimit} to 0, implying that 
\begin{equation}
    \begin{split}
        \phi^+_{L+}(r,\hat{n}) &= \int [d\Delta](2r)^{-\Delta}\alpha^+_\lambda(\hat{n}) \\
        \phi^{(1)}(u>0,\hat{n}) &= \frac{u}{2}\int [d\Delta] u^{-\Delta} \beta^+_\lambda(\hat{n}) + \lim_{r\to \infty}r\phi^+_{L+}(r,\hat{n}).
    \end{split}
\end{equation}
Similar computations in the other regions imply that the asymptotia of the reduced field near the AdS/dS boundary encode the value of the 4D field on the lightcone and null infinity. Defining the asymptotia by
\begin{equation}
\begin{split}
    \phi^+_\lambda(\zeta,\hat{n}) & \sim e^{-\Delta\zeta}\alpha^+_\lambda(\hat{n}) + e^{(\Delta-2)\zeta}\beta^+_\lambda(\hat{n}),\ \zeta\to\infty \\
    \phi^0_\lambda(\eta,\hat{n}) &\sim e^{-\Delta\eta}\alpha^0_\lambda(\hat{n}) + e^{(\Delta-2)\eta}\beta^0_\lambda(\hat{n}),\ \eta \to \infty \\
    \phi^0_\lambda(\eta,\hat{n}) &\sim e^{\Delta\eta}\tilde{\alpha}^0_\lambda(\hat{n}) + e^{(2-\Delta)\eta}\tilde{\beta}^0_\lambda(\hat{n}),\ \eta\to-\infty \\
    \phi^-_\lambda(\zeta,\hat{n}) &\sim e^{-\Delta\zeta}\alpha^-_\lambda(\hat{n}) + e^{(\Delta-2)\zeta}\beta^-_\lambda(\hat{n}),\ \zeta\to \infty,
\end{split}
\end{equation}
the values of the field on the lightcone are 
\begin{empheq}[box=\fbox]{equation}
    \begin{split}
        \phi^+_{L+}(r,\hat{n}) &= \int [d\Delta](2r)^{-\Delta}\alpha^+_\lambda(\hat{n}) \\
        \phi^0_{L+}(r,\hat{n}) &= \int [d\Delta] (2r)^{-\Delta}\alpha^0_\lambda(\hat{n}) \\
        \phi^0_{L-}(r,\hat{n}) &= \int [d\Delta](2r)^{-\Delta}\tilde{\alpha}^0_\lambda(\hat{n}) \\
        \phi^-_{L-}(r,\hat{n}) &= \int [d\Delta](2r)^{-\Delta}\alpha^-_{\lambda}(\hat{n})
    \end{split}
\end{empheq}
while the values on $\mathscr{I}^\pm$ are
\begin{empheq}[box=\fbox]{equation}\label{eq:minkscrilim}
    \begin{split}
        \phi^{(1)}(u > 0, \hat{n}) &=\frac{1}{2}\int [d\Delta]|u|^{1-\Delta}\beta^+_\lambda(\hat{n}) + \lim_{r\to\infty}r\phi^+_{L+}(r,\hat{n}) \\
        \phi^{(1)}(u < 0, \hat{n}) &= \frac{1}{2}\int [d\Delta]|u|^{1-\Delta}\beta^0_\lambda(\hat{n}) + \lim_{r\to \infty}r\phi^0_{L+}(r,\hat{n}) \\
        \phi^{(1)}(v>0,\hat{n}) &= \frac{1}{2}\int [d\Delta]|v|^{1-\Delta}\tilde{\beta}^0_\lambda(\hat{n}) + \lim_{r\to\infty}r\phi^0_{L-}(r,\hat{n}) \\
        \phi^{(1)}(v < 0, \hat{n}) &=  \frac{1}{2}\int [d\Delta]|v|^{1-\Delta}\beta^-_\lambda(\hat{n}) + \lim_{r\to\infty}r\phi^-_{L-}(r,\hat{n}).
    \end{split}
\end{empheq}
In these expressions, we have shifted $\alpha_0,\beta_0$ appropriately so that $\int [d\Delta] |u|^{1-\Delta}\beta_\lambda = 0$ at $u = 0$. 
\subsubsection{The Reduced Action}
We now detail how to reduce the massless action  to a set of actions on lower-dimensional slices of Minkowski space with a noncompact set of fields. We can separate out the full action into 5 terms:
\begin{equation}
    \begin{split}
        S &= S_+ + S_0 + S_- + S_{L+} + S_{L-} \\
        S_\pm &= -\frac{1}{2}\int_{\substack{x^2 < 0 \\ \pm x^0 > 0}} d^4x \partial^\mu\phi\partial_\mu\phi \\
        S_0 &= -\frac{1}{2}\int_{x^2>0}d^4x \partial^\mu\phi\partial_\mu\phi 
    \end{split}
\end{equation}
and $S_{L\pm}$ are terms living on the forward (past) pointing lightcone. Consider first the action inside the forward pointing lightcone. Here, we can expand
\begin{equation}
    \phi(\tau\hat{x}_+) = \int_{-\infty}^\infty [d\Delta]\tau^{-1-i\lambda}\phi^+_\lambda(\zeta,\hat{n}).
\end{equation}
Inserting this into the action gives
\begin{equation}\label{eq:timpmlact}
    \begin{split}
        S_+ &= -\frac{1}{2}\int[d\Delta_1][d\D_2]\int\tau^3d\tau d^3\hat{x}_+\sqrt{h_+}(-(\partial_\tau \tau^{-1-i\lambda_1})(\partial_\tau\tau^{-1-i\lambda_2})\phi^+_{\lambda_1}\phi^+_{\lambda_2} + \tau^{-4-i\lambda_1-i\lambda_2}\partial^a\phi^+_{\lambda_1}\partial_a\phi^+_{\lambda_2}) \\
        &= -\frac{1}{2}\int [d\D_1][d\D_2]\int d\tau\tau^{-1-i\lambda_1-i\lambda_2} \int d^3\hat{x}_+\sqrt{h_+}(\partial^a\phi^+_{\lambda_1}\partial_a\phi^+_{\lambda_2} - (1+i\lambda_1)(1+i\lambda_2)\phi^+_{\lambda_1}\phi^+_{\lambda_2}) \\
        &= -\frac{1}{2}\int[d\D]\int d^3\hat{x}_+\sqrt{h_+}(\partial^a\phi^+_\lambda\partial_a\phi^+_{-\lambda} - (1+\lambda^2)\phi^+_{\lambda}\phi^+_{-\lambda})
    \end{split}
\end{equation}
where we have used the integral $\int dx x^{-1-iy} = 2\pi\delta(y)$ for real $y$. We can use similar expressions in the exterior of the lightcone and the interior of the past lightcone to find 
\begin{equation}\label{eq:spmlact}
\begin{split}
    S_0 &= -\frac{1}{2}\int[d\Delta]\int d^3\hat{x}_0\sqrt{h_0}(\partial^a\phi^0_\lambda\partial_a\phi^0_{-\lambda} + (1+\lambda^2)\phi^0_\lambda\phi^0_{-\lambda})
\end{split}
\end{equation}
and 
\begin{equation}\label{eq:tmnmlact}
\begin{split}
    S_- &= -\frac{1}{2}\int [d\Delta] \int d^3\hat{x}_-\sqrt{h_-}(\partial^a\phi^-_\lambda\partial_a\phi^-_{-\lambda} - (1+\lambda^2)\phi^-_{\lambda}\phi^-_{-\lambda}).
\end{split}
\end{equation}
One might think that the actions in Equations \eqref{eq:timpmlact}, \eqref{eq:spmlact}, and \eqref{eq:tmnmlact} are together equivalent to the full massless action, as the lightcone is a set of measure 0. However, if fields are discontinuous across a null surface there can be a distributional term in the action on that null surface. In our case, because we treat the fields on either side of the lightcone as effectively different species of particles, there will be a $\delta$-function-like term in $(\partial\phi)^2$ that will localize the integral to the lightcone. To study terms on the future lightcone, we use the coordinate system 
\begin{equation}
   x^\mu = r (1,\hat{n}) + t(1,-\hat{n}).
\end{equation}
In these coordinates, 
\begin{equation}
    x^2 = x^\mu x_\mu = -4r t
\end{equation}
where $r > 0$ and $t \in \mathbb{R}$. The region with $t > 0$ will cover the interior of the lightcone and $t < 0$ the exterior. The metric takes the form
\begin{equation}
    ds^2 = -4dr dt + (r-t)^2d\Omega_2^2.
\end{equation}
The Lagrangian takes the form
\begin{equation}
    \mathcal{L} = -\frac{1}{2}\partial^\mu\phi\partial_\mu\phi = \frac{1}{2}\partial_r \phi\partial_t\phi -\frac{1}{2(t-r)^2}\nabla_2^a\phi\nabla_{2a}\phi
\end{equation}
where $\nabla_2$ is the connection on the unit 2-sphere. Now, suppose that the field is discontinuous across the lightcone with form 
\begin{equation}
    \phi = \Theta(t)\phi_+(r,t,\hat{n}) + \Theta(-t)\phi_0(r,t,\hat{n}).
\end{equation}
Distributional terms will arise from the derivatives in $t$ hitting the $\Theta$ functions. Hence, assuming that $\Theta(t)\delta(t) = 1/2\delta(t)$, 
\begin{equation}
    \begin{split}
        \mathcal{L} &\supset \frac{1}{4}(\phi_+(r,0,\hat{n})-\phi_0(r,0,\hat{n}))\partial_r (\phi_+(r,0,\hat{n}) + \phi_0(r,0,\hat{n}))\delta(t).
    \end{split}
\end{equation}
Hence, 
\begin{equation}
\begin{split}
        S_{L+} = \frac{1}{2}\int r^2dr d^2\hat{n}(\phi^+_{L+}-\phi^0_{L+})\partial_r(\phi^+_{L+} + \phi^0_{L+}) .
    \end{split}
\end{equation}
The action on the past-directed lightcone is 
\begin{equation}
    S_{L-} = \frac{1}{2}\int r^2dr d^2\hat{n}(\phi^-_{L-}-\phi^0_{L-})\partial_r(\phi^0_{L-} + \phi^-_{L-}).
\end{equation}
Inserting the appropriate lightcone limits of the reduced field and evaluating the $r$ integral gives
\begin{equation}
\begin{split}
    S_{L+} &= -\frac{1}{8}\int d^2\hat{n} \int [d\Delta](1-i\lambda)(\alpha^+_\lambda-\alpha^0_\lambda)(\alpha^+_{-\lambda} + \alpha^0_{-\lambda}) \\
    S_{L-} &= -\frac{1}{8}\int d^2\hat{n} \int [d\Delta](1-i\lambda)(\alpha^-_\lambda-\tilde{\alpha}^0_\lambda)(\tilde{\alpha}^0_{-\lambda}+\alpha^-_{-\lambda}).
    \end{split}
\end{equation}
Altogether, the reduced action takes the form
\begin{empheq}[box=\fbox]{equation}\label{eq:minkmasslessaction}
\begin{split}
S = &-\frac{1}{2}\int [d\Delta] \int d^3\hat{x}_+\sqrt{h_+}(\partial^a\phi^+_\lambda\partial_a\phi^+_{-\lambda}-(1+\lambda^2)\phi^+_\lambda\phi^+_{-\lambda}) \\
&- \frac{1}{2}\int [d\Delta] \int d^3\hat{x}_0\sqrt{h_0}(\partial^a\phi^0_\lambda\partial_a\phi^0_{-\lambda} + (1+\lambda^2)\phi^0_\lambda\phi^0_{-\lambda}) \\
&-\frac{1}{2}\int [d\Delta] \int d^3\hat{x}_-\sqrt{h_-}(\partial^a\phi^-_\lambda\partial_a\phi^-_{-\lambda}-(1+\lambda^2)\phi^-_\lambda\phi^-_{-\lambda}) \\
&- \frac{1}{8}\int [d\Delta]\int d^2\hat{n}(1-i\lambda)(\alpha^+_\lambda - \alpha^0_\lambda)(\alpha^+_{-\lambda}+\alpha^0_{-\lambda}) \\
&- \frac{1}{8}\int [d\Delta]\int d^2\hat{n}(1-i\lambda)(\alpha^-_\lambda-\tilde{\alpha}^0_\lambda)(\alpha^-_{-\lambda} + \tilde{\alpha}^0_{-\lambda})
\end{split}
\end{empheq}
where $\alpha^{\pm/0}$ are defined from the expansions of the reduced field near the asymptotic boundary. 
\subsubsection{Interactions}
Interaction terms can similarly be reduced. Consider the interaction term 
\begin{equation}
    S_{int} = \frac{g}{n!}\int d^4x\phi^n(x).
\end{equation}
Because this interaction term has no derivatives, there are no lightcone contributions and we can separate it into the contributions that are in the interior and exterior of the lightcone, giving 
\begin{equation}
    S_{int} = S_{int,+} + S_{int,0} + S_{int,-}.
\end{equation}
Inserting the appropriate field reduction gives
\begin{equation}
    S_{int,+} = \frac{g}{n!}\int\prod_{j=1}^n [d\Delta_j]\int \tau^{3-\sum_j\Delta_j}d\tau\int d^3\hat{x}_+\sqrt{h_+}\prod_{j=1}^n\phi^+_{\lambda_j}(\zeta,\hat{n}).
\end{equation}
The integral over $\tau$ is distributionally convergent when $\mathrm{Re}[\sum_j\Delta_j]=4$. On the locus where $\mathrm{Re}[\sum_j\Delta_j] = 4$, the interaction vertex takes the form 
\begin{equation}
    \begin{split}
        S_{int,+} 
        &= \frac{2\pi g}{n!}\int \prod_{j=1}^n [d\Delta_j]\delta(\sum_j\Delta_j-4)\int d^3\hat{x}_+\sqrt{h_+}\prod_{j=1}^n\phi^+_{\lambda_j}(\zeta,\hat{n}) \\
       S_{int,0} &=  \frac{2\pi g}{n!}\int \prod_{j=1}^n [d\Delta_j]\delta(\sum_j\Delta_j-4)\int d^3\hat{x}_0\sqrt{h_0} \prod_{j=1}^n\phi^0_{\lambda_j}(\eta,\hat{n}) \\
       S_{int,-} &= \frac{2\pi g}{n!}\int \prod_{j=1}^n [d\Delta_j]\delta(\sum_j\Delta_j-4)\int d^3\hat{x}_-\sqrt{h_-}\prod_{j=1}^n\phi^-_{\lambda_j}(\zeta,\hat{n}).
    \end{split}
\end{equation}
For $n = 4$, this is satisfied when all external weights are on the continuous principal series, which is due to the marginality of the operator $\phi^4$ in 4 dimensions.  For $n \ne 4$, the integral can be defined by continuing away from the principal series. These operators can be interpreted as contour integrals along the continuous principal series \cite{Donnay:2018neh}. This divergence is reminiscent of the divergent center-of-momentum integral for Mellin transforms of scattering amplitudes for tree-level massless scattering \cite{Puhm:2019zbl, Arkani-Hamed:2020gyp, Borji:2024pvg}. 
\subsection{Reconstructing 4D Physics}
In this section, we show how to extract four-dimensional physics from these lower dimensional actions. First, we understand how the lightcone terms in the action lead to propagation across the lightcone. We then show that this correctly reproduces the equations of motion for a spherically symmetric solution. Finally, we argue that the reduced action generates the free two-point function. 
\subsubsection{Propagation Across the Lightcone}
In this section, we find the equation of motion for the action in Equation \eqref{eq:minkmasslessaction} and show how it reproduces free propagation across the lightcone. 

To extract the equations of motion, we vary the fields by  
\begin{equation}
    \phi_\l^a \to \phi_\l^a + \delta\phi_\l^a
\end{equation}
Then, in the interior of each region the variation of the action is 
\begin{equation}
    \begin{split}
        \delta S_+ &= -\int [d\D]\int d^3\hat{x}_+\sqrt{h_+}(\del^a\phi_\l^+\del_a\delta\phi^+_{-\l} - (1+\l^2)\phi^+_\l\delta\phi^+_{-\l}) \\
        &= -\frac{1}{4}\int[d\D]\int_{\partial H_3}d^2\hat{n}e^{2\zeta}\partial_\zeta\phi^+_\l\delta\phi^+_{-\l} + \int[d\D]\int d^3\hat{x}_+\sqrt{h_+}(\square_{H_3}\phi^+_\l+(1+\l^2)\phi^+_\l)\delta\phi^+_{-\l}
    \end{split}
\end{equation}
Similarly, 
\begin{equation}
    \delta S_- = -\frac{1}{4}\lim_{\zeta\to\infty}\int[d\D]\int d^2\hat{n}e^{2\zeta}\del_\zeta\phi^-_\l\delta\phi^-_{-\l} + \int [d\D]\int d^3\hat{x}_-\sqrt{h_-}(\square_{H_3}\phi^-_\l+(1+\l^2)\phi^-_\l)\delta\phi^-_{-\l}
\end{equation}
In the region $x^2 > 0$, 
\begin{equation}
\begin{split}
    \delta S_0 &= \int [d\D]\int d^3\hat{x}_0\sqrt{|h_0|}(\square_{dS_3}\phi^0_\l-(1+\l^2)\phi^0_\l)\delta\phi^0_{-\l} \\
    &+ \lim_{R\to\infty}\frac{1}{4}\int [d\D]\int d^2\hat{n}e^{2|\eta|}\del_\eta\phi^0_\l\delta\phi^0_{-\l}|^{\eta=R}_{\eta=-R}
\end{split}
\end{equation}
Hence, the reduced action reproduces the free equations of motion for the reduced field. 

To see how fields propagate across the lightcone, we have to include the boundary contributions from $S_{\pm,0}$ that arise from integrating by parts as well as the lightcone terms. Inserting the asymptotic expansion of the fields, we have that\footnote{We have dropped highly oscillatory terms when taking the boundary limit. These vanish for sufficiently regular fields.}
\begin{equation}
    \begin{split}
        (\delta S_+ + \delta S_- + \delta S_0)_{bdry} &= \frac{1}{4}\int [d\D]\int d^2\hat{n}\left[(1+i\l)(\alpha^+_\l\delta\alpha^+_{-\l}+\alpha^-_\l\delta\alpha^-_{-\l} - \alpha^0_{\l}\delta\alpha^0_{-\l}-\tilde{\alpha}^0_\l\delta\tilde{\alpha}^0_{-\l} \right. \\
        &\left.+(1-i\l)(\beta^+_\l\delta\beta^+_{-\l} +\beta^-_{\l}\delta\beta^{-}_{-\l}-\beta^0_{\l}\delta\beta^0_{-\l}-\tilde{\beta}^0_\l\delta\tilde{\beta}^0_{-\l}\right].
    \end{split}
\end{equation}
We also have to vary the explicit lightcone terms, which give
\begin{equation}
    \begin{split}
        \delta S_{L+} &= -\frac{1}{8}\int [d\D]\int d^2\hat{n}(1-i\l)[(\delta\alpha^+_\l-\delta\alpha^0_{\l})(\alpha^+_{-\l}+\alpha^0_{-\l}) + (\alpha^+_{\l}-\alpha^0_{\l})(\delta\alpha^+_{-\l}+\delta\alpha^0_{-\l})] \\
        &= -\frac{1}{4}\int [d\D]\int d^2\hat{n}[(\alpha^+_\l+i\l\alpha^0_\l)\delta\alpha^+_{-\l}-(\alpha^0_\l+i\l\alpha_\l^+)\delta\alpha^0_{-\l}] \\
        \delta S_{L-} &= -\frac{1}{4}\int[d\D]\int d^2\hat{n}[(\alpha^-_\l+i\l\tilde{\alpha}^0_\l)\delta\alpha^-_{-\l}-(\tilde{\alpha}^0_\l+i\l\alpha^-_\l)\delta\tilde{\alpha}^0_{-\l}].
    \end{split}
\end{equation}
Combining all of the lightcone terms and setting $\delta\beta^a = \delta\tilde{\beta}^a = 0$ so that the value of the field on null infinity is fixed gives
\begin{equation}
\begin{split}
\delta S_{bdry} &=  (\delta S_+ + \delta S_- + \delta S_0)_{bdry} + \delta S_{L+} + \delta S_{L-} \\
&=\frac{i}{4}\int[d\D]\int d^2\hat{n}\l[(\alpha^+_\l-\alpha^0_\l)(\delta\alpha^+_{-\l} + \delta\alpha^0_{-\l}) + (\alpha^-_\l - \tilde{\alpha}^0_{\l})(\delta\alpha^-_{-\l} + \delta\tilde{\alpha}^0_{-\l})]
\end{split}
\end{equation}
Hence, requiring that this action be stationary sets
\begin{equation}
    \alpha^+_\l = \alpha^0_\l,\ \tilde{\alpha}^0_\l = \alpha^-_\l,\ \l \ne 0.
\end{equation}
Hence, for modes with $\l \ne 0$, we simply have continuity across the lightcone. For modes with $\lambda = 0$, solutions with $\alpha^+_0 \ne \alpha^0_0$ correspond to radially propagating shockwaves which are singular at $r = 0$. Requiring solutions be regular everywhere requires $\alpha^+_0 = \alpha^0_0$ and $\tilde{\alpha}^0_0 = \alpha^-_0$. 

We now show that this reproduces the propagation of spherically symmetric solutions of the massless Klein-Gordon equation. A general spherically-symmetric solution to the massless Klein-Gordon equation in Minkowski space can be written 
\begin{equation}
    \phi(t,r) = \frac{F(t+r) - F(t-r)}{r}
\end{equation}
Taking the limit as $(t,r) \to \scri^\pm$ gives
\begin{equation}
    \begin{split}
        \phi^{(1)}(v) = \lim_{\substack{r\to\infty\\ t+r=v}} r\phi = F(v) - F(-\infty) \\
        \phi^{(1)}(u) = \lim_{\substack{r\to\infty\\ t-r=u}} r\phi = F(\infty)-F(u).
    \end{split}
\end{equation} Consider for simplicity a solution that obeys $F(v) = 0$ for $v \le 0$ and $F(v) \to 0$ for $v\to\infty$. Then, we must have that the reduced field satisfies
\begin{equation}
    \begin{split}
        \phi^{(1)}(v) &=\frac{1}{2}\int[d\Delta]|v|^{1-\Delta}\tilde{\beta}^0_\l \\
        \tilde{\beta}^0_\l &= 2\int_0^\infty dvv^{i\l-1}F(v) \\
        \tilde{\alpha}^0_\l &= 0
    \end{split}
\end{equation}
Hence, the reduced spatial field obeys
\begin{equation}
    \phi^0_\l(\eta,\hat{n}) \sim e^{(2-\D)\eta}\tilde{\beta}^0_\l\ \mathrm{as\ } \eta\to-\infty
\end{equation}
Using the equation of motion for a free field in dS$_3$, we then have that 
\begin{equation}
\phi^0_\l(\eta,\hat{n})\sim e^{-\Delta\eta}\tilde{\beta}^0_{\lambda}\ \mathrm{as\ } \eta\to\infty.
\end{equation}
Inserting this value into our reconstruction formula implies that 
\begin{equation}
\begin{split}
    \phi^{(1)}(u < 0,\hat{n}) &= 0 \\
    \phi^0_{L+}(r,\hat{n}) &= 2\int [d\Delta](2r)^{-\Delta}\int_0^\infty dvv^{\D-2}F(v) = \frac{F(2r)}{r}
\end{split}
\end{equation}
which correctly reproduces the value of the field on the future-directed lightcone. The equations of motion then require
\begin{equation}
    \alpha^+_\l = \alpha^0_\l = \tilde{\beta}^0_\l.
\end{equation}
For the resulting solution to be nonsingular on the AdS$_3$ slice, we must have that
\begin{equation}
    \beta^+_\l = -\alpha^+_\l = -\tilde{\beta}^0_\l
\end{equation}

Hence, we have that the reduced equations of motion and the boundary term imply that 
\begin{equation}
    \phi^{(1)}(u) = -\int [d\Delta]u^{1-\D}\int_0^\infty dvv^{\D-2}F(v) = -F(u)
\end{equation}
This agrees with the calculation using the 4D action. 
\subsubsection{Two-Point Functions}
In this section we argue that the action in Equation \eqref{eq:minkmasslessaction} appropriately reproduces the two-point function. Consider the partition function 
\begin{equation}
\begin{split}
    Z[J^+_\lambda(\zeta,\hat{n}), J^0_\lambda(\eta,\hat{n}), J^-_\lambda(\zeta,\hat{n})] &= \int D\phi^{\pm,0}_\lambda e^{i S + i\int [d\Delta] \int d^3\hat{x}_+\sqrt{h_+}J^+_\lambda\phi^+_\lambda} \\
    &\times e^{i\int [d\Delta]\int d^3\hat{x}_0\sqrt{h_0}J^0_\lambda \phi^0_\lambda + i\int[d\Delta]\int d^3\hat{x}_-\sqrt{h_-}J^-_\lambda\phi^-_\lambda}.
\end{split}
\end{equation}
Correlation functions of $\phi^{\pm/0}_\l$ can be extracted by taking functional derivatives of this partition function with respect to $J^{\pm,0}_\lambda$. 

Letting 
\begin{equation}
    J(x) = \begin{cases}
        \int [d\Delta] \tau^{\Delta-4}J^+_\lambda(\zeta,\hat{n}) & x^2 < 0,\ x^0 > 0 \\
        \int [d\Delta] \rho^{\Delta-4}J^0_\lambda(\eta,\hat{n}) & x^2 > 0 \\
        \int [d\Delta] \tilde{\tau}^{\Delta-4}J^-_\lambda(\zeta,\hat{n}) & x^2 < 0,\ x^0 < 0 
    \end{cases},
\end{equation}
the source term becomes 
\begin{equation}
    i\int [d\Delta] \int d^3\hat{x}_+\sqrt{h_+}J^+_\lambda\phi^+_\lambda + i\int [d\Delta]\int d^3\hat{x}_0\sqrt{h_0}J^0_\lambda \phi^0_\lambda + i\int[d\Delta]\int d^3\hat{x}_-\sqrt{h_-}J^-_\lambda\phi^-_\lambda = i\int d^4x J(x)\phi(x).
\end{equation}
Using the relations 
\begin{equation}
\begin{split}
    \int [d\Delta]\tau^{-\Delta}\frac{\delta}{\delta J_\lambda^+(\zeta,\hat{n})} &= \frac{\delta}{\delta J(\tau\hat{x}_+)} \\
    \int [d\Delta]\rho^{-\Delta}\frac{\delta}{\delta J_\lambda^0(\eta,\hat{n})} &= \frac{\delta}{\delta J(\rho\hat{x}_0)} \\
    \int [d\Delta]\tilde{\tau}^{-\Delta}\frac{\delta}{\delta J_\lambda^-(\zeta,\hat{n})} &= \frac{\delta}{\delta J(-\tilde{\tau}\hat{x}_-)}
\end{split}
\end{equation}
we see that 
\begin{equation}
    \begin{split}
        \int [d\Delta_j]\tau_1^{-\Delta_1}\tau_2^{-\Delta_2}\langle \phi^+_{\lambda_1}(\zeta_1,\hat{n}_1)\phi^+_{\lambda_2}(\zeta_2,\hat{n}_2)\rangle &= -\int [d\Delta_j]\tau_1^{-\Delta_1}\tau_2^{-\Delta_2} \frac{\delta^2 Z}{\delta J^+_{\lambda_1}(\zeta_1,\hat{n}_1)\delta J^+_{\lambda_2}(\zeta_2,\hat{n}_2)}|_{J = 0} \\
        &= -\frac{\delta^2Z}{\delta J(\tau_1\hat{x}_+(\zeta_1,\hat{n}_1))\delta J(\tau_2\hat{x}_+(\zeta_2,\hat{n}_2))}|_{J=0} \\
        &= \langle \phi(\tau_1\hat{x}_+(\zeta_1,\hat{n}_1))\phi(\tau_2\hat{x}_+(\zeta_2,\hat{n}_2))\rangle.
    \end{split}
\end{equation}
Identical considerations imply that the two-point functions are correctly reproduced in all regions of spacetime. Note that the lightcone contribution to the action is essential here; without it, the two-point function $\langle \phi^+_\l(\zeta,\hat{n})\phi^0_{\l'}(\eta',\hat{n}')\rangle$ would factorize into a product of one-point functions. 
\subsection{Klein Space}
The structure of the action in Klein space is somewhat simpler as the lightcone is fully connected. In this region, we can reduce the bulk field by letting 
\begin{equation}
    \phi(\tau\hat{x}_+) = \int [d\Delta]\tau^{-\Delta}\phi^+_\lambda(\rho,\psi,\phi)
\end{equation}
inside the lightcone and 
\begin{equation}
    \phi(\tau\hat{x}_-) = \int [d\Delta]\tau^{-\Delta}\phi^-_\lambda(\rho,\psi,\phi)
\end{equation}
outside it.
\subsubsection{Boundary Limits}
At large values of $\rho$, the reduced field can be expanded as 
\begin{equation}
\phi^\pm_\lambda(\rho,\psi,\phi) = e^{-\Delta\rho}\alpha^\pm_\lambda(\psi,\phi) + e^{(\Delta-2)\rho}\beta^\pm_\lambda(\psi,\phi). 
\end{equation}
The lightcone is reached from either region by sending $\tau\to 0, \rho \to \infty$ while fixing $\tau e^\rho = 2r$. In this limit, we have that 
\begin{equation}
    \begin{split}
        \phi^+_L(r,\psi,\phi) &= \lim_{\substack{\rho\to\infty \\ \tau e^\rho = 2r}}\int [d\Delta]\tau^{-\Delta}\phi^+_\lambda(\rho,\psi,\phi) \\
        &= \lim_{\rho\to\infty}\int [d\Delta](2r e^{-\rho})^{-\Delta}\phi^+_\lambda(\rho,\psi,\phi) \\
        &= \lim_{\rho\to\infty} \int [d\Delta] (2r)^{-\Delta} [\alpha^+_\lambda(\psi,\phi) + e^{2(\Delta-1)\rho}\beta^+_\lambda(\psi,\phi)] \\
        &= \int [d\Delta](2r)^{-\Delta}\alpha^+_\lambda(\psi,\phi).
    \end{split}
\end{equation}
We can again choose the split between $\alpha^+_0$ and $\beta^+_0$ so that the latter term does not contribute. 

From the spacelike region, we have that 
\begin{equation}
    \begin{split}
        \phi^-_L(r,\psi,\phi) &=  \lim_{\substack{\rho\to\infty \\ \tau e^\rho = 2r}}\int [d\Delta]\tau^{-\Delta}\phi^-_\lambda(\rho,\phi,\psi) \\
         &=  \int [d\Delta] (2r)^{-\Delta} [\alpha^-_\lambda(\phi,\psi)]
    \end{split}
\end{equation}
recalling the flip between the two angles in Equations \eqref{eq:kslclt} and \eqref{eq:kslcls} and defining the split between $\alpha_0$ and $\beta_0$ so that the $\beta$ term vanishes.

The region of $\scri$ with $u > 0$ is reached from the interior of the lightcone by sending $\rho\to\infty$ while fixing $u = \tau e^{-\rho}$. In this limit,
\begin{equation}
    \begin{split}
        \phi^{(1)}(u>0,\psi,\phi) &= \lim_{\substack{\rho\to\infty \\ \tau e^{-\rho} = u}} \frac{\tau e^\rho}{2}\int [d\Delta]\tau^{-\Delta}\phi^+_\lambda(\rho,\psi,\phi) \\
        &= \frac{u}{2}\int [d\Delta] u^{-\Delta}\beta^+_\lambda(\psi,\phi) + \lim_{r\to\infty}r \phi^+_L(r,\psi,\phi).
    \end{split}
\end{equation}
The other half of $\mathscr{I}$ is reached from the exterior of the lightcone by sending $\rho \to \infty$ while fixing $u = -\tau e^{-\rho}$. In this limit, then, we have that 
\begin{equation}
    \begin{split}
        \phi^{(1)}(u < 0, \psi,\phi) &= \lim_{\substack{\rho\to\infty \\ \tau e^{-\rho} = -u}}\frac{\tau e^\rho}{2}\int [d\Delta]\tau^{-\Delta}\phi^-_\lambda(\rho,\phi,\psi) \\
        &= -\frac{u}{2}\int [d\Delta] (-u)^{-\Delta}\beta^-_\lambda(\phi,\psi) + \lim_{r\to\infty} r\phi^-_L(r,\psi,\phi).
    \end{split}
\end{equation}
In summary, the asymptotia of the reduced field encode the values on the lightcone as
\begin{empheq}[box=\fbox]{equation}
\begin{split}
    \phi^+_L(r,\psi,\phi) &=  \int [d\Delta](2r)^{-\Delta}\alpha^+_\lambda(\psi,\phi) \\
    \phi^-_L(r,\psi,\phi) &= \int [d\Delta] (2r)^{-\Delta} \alpha^-_\lambda(\phi,\psi)
\end{split}
\end{empheq}
and on null infinity as 
\begin{empheq}[box=\fbox]{equation}
    \begin{split}
        \phi^{(1)}(u>0,\psi,\phi) &= \frac{|u|}{2}\int [d\Delta] |u|^{-\Delta}\beta^+_\lambda(\psi,\phi) + \lim_{r\to\infty}r \phi^+_L(r,\psi,\phi) \\
        \phi^{(1)}(u<0,\psi,\phi) &= \frac{|u|}{2}\int [d\Delta] |u|^{-\Delta}\beta^-_\lambda(\phi,\psi) + \lim_{r\to\infty} r\phi^-_L(r,\psi,\phi).
    \end{split}
\end{empheq}
\subsubsection{The Reduced Action}
We can now evaluate the reduced action, which will come in three parts:
\begin{equation}
    \begin{split}
        S &= -\frac{1}{2}\int d^4x\partial^\mu\phi\partial_\mu\phi = S_+ +S_- + S_L \\
        S_\pm &= -\frac{1}{2}\int_{\pm x^2 < 0}d^4x\partial^\mu\phi\partial_\mu\phi.
    \end{split}
\end{equation}
We have that 
\begin{equation}
    \begin{split}
        S_+ &= -\frac{1}{2}\int_{x^2 < 0}d^4x\partial^\mu\phi\partial_\mu\phi \\
        &= -\frac{1}{2}\int \tau^3d\tau d^3\hat{x}\sqrt{h}(-(\partial_\tau\phi)^2+\frac{1}{\tau^2}\partial^a\phi\partial_a\phi) \\
        &= -\frac{1}{2}\int[d\Delta]\int d^3\hat{x}\sqrt{h}(\partial^a\phi^+_\lambda\partial_a\phi^+_{-\lambda} - (1+\lambda^2)\phi^+_\lambda\phi^+_{-\lambda})
    \end{split}
\end{equation}
while 
\begin{equation}
    S_- = \frac{1}{2}\int[d\Delta]\int d^3\hat{x}\sqrt{h}(\partial^a\phi^-_\lambda\partial_a\phi^-_{-\lambda} - (1+\lambda^2)\phi^-_\lambda\phi^-_{-\lambda}).
\end{equation}
We can similarly extract the lightcone term, as above, giving 
\begin{equation}
    S_L = \frac{1}{2}\int r^2dr d\psi d\phi(\phi^+_L-\phi^-_L)\partial_r(\phi^+_L+\phi^-_L).
\end{equation}
Inserting the expansions of the field on the lightcone gives
\begin{equation}
S_L = -\frac{1}{8}\int [d\Delta](1-i\lambda)\int d\phi d\psi (\alpha^+_\lambda(\psi,\phi)-\alpha^-_\lambda(\phi,\psi))(\alpha^+_{-\lambda}(\psi,\phi)+\alpha^-_{-\lambda}(\phi,\psi)).
\end{equation}
The full reduced action is therefore
\begin{empheq}[box=\fbox]{equation}\label{eq:kleinact}
\begin{split}
S &= -\frac{1}{2}\int[d\Delta]\int d^3\hat{x}\sqrt{h}(\partial^a\phi^+_\lambda\partial_a\phi^+_{-\lambda} - (1+\lambda^2)\phi^+_\lambda\phi^+_{-\lambda} - \partial^a\phi^-_\lambda\partial_a\phi^-_{-\lambda} + (1+\lambda^2)\phi^-_\lambda\phi^-_{-\lambda}) \\
&- \frac{1}{8}\int [d\Delta](1-i\lambda)\int d\phi d\psi (\alpha^+_\lambda(\psi,\phi)-\alpha^-_\lambda(\phi,\psi))(\alpha^+_{-\lambda}(\psi,\phi)+\alpha^-_{-\lambda}(\phi,\psi)).
\end{split}
\end{empheq}
This action reproduces several of the features that were seen to be essential for quantizing fields on AdS$_3/\mathbb{Z}$: the fields have mass exclusively below the Breitenlohner-Freedman bound, allowing a non-negative inner product on a set of periodic solutions to be defined, and have an off-diagonal kinetic term \cite{Melton:2025ecj}.

\section{Massive Scalars}\label{sec:massive}
We can perform a similar reduction for free massive particles in Minkowski space, with action 
\begin{equation}
    S = -\frac{1}{2}\int d^4x(\partial^\mu\phi\partial_\mu\phi + m^2\phi^2).
\end{equation}

\subsection{Field Reductions}

The reduction to a lower-dimensional slice is more complicated for a massive particle. For simplicity, we start by performing the reduction in the spacelike region $x^2 > 0$. Here, we want to set 
\begin{equation}
    \phi(\rho\hat{x}_0) = \int [d\Delta] R_\lambda(\rho) \phi^0_\lambda(\eta,\hat{n})
\end{equation}
in such a way that $\square\phi = m^2\phi$ is equivalent to $\square_{dS_3}\phi^0_\lambda = (1+\lambda^2)\phi^0_\lambda$. Inserting the above expansion, this requires that 
\begin{equation}
    \rho^2 R_\lambda''(\rho) + 3\rho R_\lambda'(\rho) +(1+\lambda^2-m^2\rho^2)R_\lambda(\rho) = 0.
\end{equation}
The general solution to this equation is
\begin{equation}
    R_\lambda(\rho) = \frac{\alpha K_{i\lambda}(m\rho) + \beta I_{i\lambda}(m\rho)}{\rho}.
\end{equation}
At large $\rho$, these behave as 
\begin{equation}
    R_\lambda(\rho) \sim \rho^{-3/2}(\alpha e^{-m\rho} + \beta e^{m\rho}).
\end{equation}
Because we expect massive fields to fall off rather than exponentially grow near spacelike infinity, we throw away the second solution, implying that the proper field reduction in the spacelike region is 
\begin{equation}
    \phi(\rho\hat{x}_0) = \int [d\Delta]_+R_\lambda(\rho)\phi^0_\lambda(\eta,\hat{n}),\ R_\lambda(\rho) = \frac{K_{i\lambda}(m\rho)}{\rho}
\end{equation}
where $\int [d\D]_+ = \int_0^\infty \frac{d\lambda}{2\pi}$, and we integrate only over $\lambda > 0$ since $R_\l = R_{-\l}$.

In the interior of the future-directed lightcone, we have that
\begin{equation}
    ds^2 = -d\tau^2 + \tau^2(d\zeta^2+\sinh^2\zeta d\Omega_2^2).
\end{equation}
We now want to expand the field as 
\begin{equation}
    \phi(\tau\hat{x}_+) = \int [d\Delta] T_\lambda(\tau) \phi_\lambda^+(\zeta,\hat{n})
\end{equation}
where 
\begin{equation}
    \square \phi = m^2\phi \implies \square_{H_3}\phi^+_\lambda = -(1+\lambda^2)\phi^+_\lambda.
\end{equation}
This implies that 
\begin{equation}
    \tau^2 T_\lambda''(\tau) + 3\tau T_\lambda'(\tau) + (1 + \lambda^2 + m^2\tau^2)T_\lambda(\tau) = 0.
\end{equation}
The general solution is then 
\begin{equation}
    T_\lambda(\tau) = \frac{\alpha H^{(1)}_{i\lambda}(m\tau) + \beta H^{(2)}_{i\lambda}(m\tau)}{\tau},\ \lambda > 0.
\end{equation}
Now, the solutions asymptote to $e^{\pm im\tau}/\tau^{3/2}$, and no linear combination can be discarded. A convenient choice for $T_\l$ is
\begin{equation}
    T_\lambda(\tau) = \frac{1}{\tau} J_{-i\lambda}(m\tau),\ \lambda \in \mathbb{R}
\end{equation}
so that our reduction formula is 
\begin{equation}
    \phi(\tau\hat{x}_+) = \int [d\Delta]T_\lambda(\tau)\phi^+_\lambda(\zeta,\hat{n}),\ T_\l(\tau) = \frac{J_{-i\l}(m\tau)}{\tau}
\end{equation}
where $\lambda \in \mathbb{R}$. Similarly, we can expand the field in the interior of the past lightcone by letting 
\begin{equation}
    \phi(\tau\hat{x}_-) = \int [d\Delta] T_\lambda(-\tau)\phi^-_\lambda(\zeta,\hat{n}).
\end{equation}
Because $J_{i\l} \ne J_{-i\l}$, we need to integrate over the entire $\lambda$ line. While this may seem inconsistent with the spacelike answer, where half of the radial profiles could be discarded by applying appropriate boundary conditions, we have to keep both so that the asymptotic behavior of the field at large $\tau$ contains both positive and negative frequency parts. Alternatively, we could choose to define our timelike profile by analytic continuation from the spacelike region; in this case, we would get a doubling of the number of reduced fields by continuing $\rho \to i\tau$ or $\rho \to -i\tau$, which generate positive and negative frequencies at $i^\pm$ respectively. Choosing $T_\l(\tau) = J_{-i\l}(m\tau)/\tau$ is convenient for studying the action near the lightcone. 
\subsection{Boundary Limits}
Using our reductions, we have that
\begin{equation}
    \begin{split}
        \phi^+_{L+} &= \lim_{\substack{\zeta\to\infty \\ \tau e^{\zeta} = 2r}}\int [d\Delta]\frac{J_{-i\lambda}(m\tau)}{\tau}\phi^+_{\lambda}(\zeta,\hat{n}) \\
        &= \lim_{\zeta\to \infty}\int [d\Delta]\frac{J_{-i\lambda}(2mr e^{-\zeta})}{2r e^{-\zeta}}\phi^+_\lambda(\zeta,\hat{n}) \\
        &= \lim_{\zeta\to \infty} \int [d\Delta]\left(\frac{m}{2}\right)^{1-\Delta}\frac{1}{\G(2-\D)}(2r e^{-\zeta})^{-\D}\phi^+_\lambda(\zeta,\hat{n}) \\
        &= \lim_{\zeta\to \infty}\frac{1}{2}\int [d\Delta] \frac{m^{1-\Delta}}{\G(2-\D)}r^{-\Delta}[\alpha^+_\lambda(\hat{n}) + e^{2(\Delta-1)\zeta}\beta^+_\lambda(\hat{n})] \\
        &= \int [d\Delta]\frac{m}{2\G(2-\D)}(mr)^{-\D}\alpha^+_\l(\hat{n}).
    \end{split}
\end{equation}
Similarly, we have that 
\begin{equation}
    \phi^-_{L-} = \int [d\D]\frac{m}{2\G(2-\D)}(mr)^{-\D}\alpha^-_\l(\hat{n}).
\end{equation}
The exterior of the lightcone is more complicated due to our restriction that $R_\lambda$ must fall off at large $\rho$. At small $\rho$, 
\begin{equation}
    \frac{K_{\D-1}(m\rho)}{\rho} \sim \frac{m^{1-\D}\G(\D-1)}{2^{2-\D}}\rho^{-\D} + \frac{m^{\D-1}\G(1-\D)}{2^\D}\rho^{\D-2}.
\end{equation}
Asymptotically, $\phi^0_\lambda$ approaches 
\begin{equation}
    \phi^0_\lambda(\eta,\hat{n}) \sim \begin{cases} 
        e^{-\D\eta}\alpha^0_\lambda(\hat{n}) + e^{(\D-2)\eta}\beta^0_\lambda(\hat{n}) & \eta \to \infty \\
        e^{\D\eta}\tilde{\alpha}^0_\lambda(\hat{n}) + e^{(2-\D)\eta}\tilde{\beta}^0_\lambda(\hat{n}) & \eta \to -\infty 
    \end{cases}.
\end{equation}
As we approach the future-pointing lightcone, we have that $\eta \to \infty$ while fixing $\rho e^\eta = 2r$. In this limit,
\begin{equation}
    \begin{split}
        \phi^0_{L+} &= \lim_{\substack{\eta\to\infty \\ \rho e^\eta = 2r}}\int [d\Delta]_+ \frac{K_{\D-1}(m\rho)}{\rho}\phi^0_\lambda(\eta,\hat{n}) \\
        &= \lim_{\eta\to\infty}\int [d\Delta]_+\frac{K_{\D-1}(2m r e^{-\eta})}{2r e^{-\eta}}\phi^0_\lambda(\eta,\hat{n}).
    \end{split}
\end{equation}
Inserting the appropriate limits gives
\begin{equation}
    \begin{split}
         \phi^0_{L+} &= \lim_{\eta\to\infty}\int [d\Delta]_+\left[\left( \frac{m^{1-\D}\G(\D-1)}{2^{2-\D}}(2r e^{-\eta})^{-\D} + \frac{m^{\D-1}\G(1-\D)}{2^\D}(2r e^{-\eta})^{\D-2}\right)\right. \\
         &\hspace{1in}\left.\left( e^{-\D\eta}\alpha^0_\lambda(\hat{n}) + e^{(\D-2)\eta}\beta^0_\lambda(\hat{n})\right)\right] \\
         &= \frac{1}{4}\int [d\Delta]_+\left[m^{1-\D}\G(\D-1)(r)^{-\D}\alpha^0_\lambda + m^{\D-1}\G(1-\D)(r)^{\D-2}\beta^0_\lambda\right] \\
         &+ \lim_{\eta\to\infty}\frac{1}{4}\int [d\Delta]_+\left[m^{1-\D}\G(\D-1)(r)^{-\D}e^{2(\D-1)\eta}\beta^0_\lambda + m^{\D-1}\G(1-\D)(r)^{\D-2}e^{2(1-\D)\eta}\alpha^0_\lambda\right].
    \end{split}
\end{equation}
The oscillatory term can contribute only when $\lambda = 0$; otherwise, the contribution would vanish in the $\eta \to \infty$ limit. Such a term would necessarily fall off as $1/r$, and could not represent a finite-energy contribution for a massive field. We therefore set it to 0, obtaining 
\begin{equation}
    \phi^0_{L+} =  \frac{1}{4}\int [d\Delta]_+\left[m^{1-\D}\G(\D-1)(r)^{-\D}\alpha^0_\lambda + m^{\D-1}\G(1-\D)(r)^{\D-2}\beta^0_\lambda\right].
\end{equation}
Similarly, we obtain 
\begin{equation}
    \phi^0_{L-} = \frac{1}{4}\int [d\Delta]_+\left[m^{1-\D}\G(\D-1)(r)^{-\D}\tilde{\alpha}^0_\lambda + m^{\D-1}\G(1-\D)(r)^{\D-2}\tilde{\beta}^0_\lambda\right].
\end{equation}
In summary, the values of the massive field on the lightcone are 
\begin{empheq}[box=\fbox]{equation}
\begin{split}
  \phi^+_{L+} &=  \int [d\Delta]\frac{m}{2\G(2-\D)}(mr)^{-\D}\alpha^+_\l(\hat{n}) \\
  \phi^0_{L+} &= \frac{1}{4}\int [d\Delta]_+\left[m^{1-\D}\G(\D-1)(r)^{-\D}\alpha^0_\lambda + m^{\D-1}\G(1-\D)(r)^{\D-2}\beta^0_\lambda\right]\\
  \phi^0_{L-} &= \frac{1}{4}\int [d\Delta]_+\left[m^{1-\D}\G(\D-1)(r)^{-\D}\tilde{\alpha}^0_\lambda + m^{\D-1}\G(1-\D)(r)^{\D-2}\tilde{\beta}^0_\lambda\right]\\
  \phi^-_{L-} &=  \int [d\D]\frac{m}{2\G(2-\D)}(mr)^{-\D}\alpha^-_\l(\hat{n}).
\end{split}
\end{empheq}
\subsection{The Reduced Action}
We now evaluate the different terms in the reduced action. Starting with the contribution from the region $x^2 > 0$, we have that
\begin{equation}
    \begin{split}
        S_0 &= -\frac{1}{2}\int_{x^2>0}d^4x[\partial^\mu\phi\partial_\mu\phi + m^2\phi^2] \\
        &= -\frac{1}{2}\int \rho^3d\rho d^3\hat{x}\sqrt{h}[\partial^\mu\phi\partial_\mu\phi + m^2\phi^2] \\
        &= -\frac{1}{2}\int\rho^3d\rho \int [d\Delta_1]_+[d\Delta_2]_+\int d^3\hat{x}\sqrt{h}\left[\frac{1}{\rho^2}R_{\lambda_1}(\rho)R_{\lambda_2}(\rho)\partial^a\phi^0_{\lambda_1}\partial_a\phi^0_{\lambda_2}  \right. \\
        &\left. +(R_{\lambda_1}'(\rho)R_{\lambda_2}'(\rho) + m^2R_{\lambda_1}(\rho)R_{\lambda_2}(\rho))\phi^0_{\lambda_1}\phi^0_{\lambda_2}\right].
    \end{split}
\end{equation}
The reduced kinetic term can be computed simply through the orthogonality of Macdonald functions given by 
\begin{equation}
    \begin{split}
        \int \rho d\rho R_{\lambda_1}(\rho)R_{\lambda_2}(\rho) = \int_0^\infty\frac{d\rho}{\rho}K_{i\lambda_1}(m\rho)K_{i\lambda_2}(m\rho)= \frac{\pi^2}{2\lambda_1\sinh\pi\lambda_1}\delta(\lambda_1-\lambda_2).
    \end{split}
\end{equation}
The second term reduces to a similar $\delta$ function integral:
\begin{equation}
    \begin{split}
        \int \rho^3d\rho [R_{\lambda_1}'(\rho)R_{\lambda_2}'(\rho) + m^2R_{\lambda_1}(\rho)R_{\lambda_2}(\rho)] &= (1+\lambda_1^2)\int_0^\infty \rho d\rho R_{\lambda_1}(\rho) R_{\lambda_2}(\rho) = \frac{\pi^2(1+\lambda_1^2)}{2\lambda_1\sinh\pi\lambda_1}\delta(\lambda_1-\lambda_2).
    \end{split}
\end{equation}
Inserting this into the action gives
\begin{equation}
\begin{split}
    S_0 &= -\frac{1}{2}\int [d\Delta]_+\frac{\pi}{4\lambda\sinh\pi\lambda}\int d^3\hat{x}\sqrt{h}[\partial^a\phi^0_\lambda\partial_a\phi^0_\lambda + (1+\lambda^2)\phi^0_\lambda\phi^0_\lambda].
\end{split}
\end{equation}

In the interior of the future-pointing lightcone, we have that 
\begin{equation}
    \begin{split}
        S_+ &= -\frac{1}{2}\int \tau^3d\tau d^3\hat{x}_+\sqrt{h_+}(-\partial_\tau\phi\partial_\tau\phi + \frac{1}{\tau^2}\partial^a\phi\partial_a\phi + m^2\phi^2) \\
        &= -\frac{1}{2}\int \tau^3d\tau \int [d\Delta_1][d\Delta_2]\int d^3\hat{x}_+\sqrt{h_+}\left(\frac{T_{\lambda_1}(\tau)T_{\lambda_2}(\tau)}{\tau^2}\partial^a\phi^+_{\l_1}\partial_a\phi^+_{\l_2} \right. \\
        &\left.+ [m^2 T_{\l_1}(\tau)T_{\lambda_2}(\tau)-T_{\l_1}'(\tau)T_{\l_2}'(\tau)]\phi^+_{\l_1}\phi^+_{\l_2}\right) \\
        &= -\frac{1}{2}\int [d\Delta_1][d\Delta_2] M(\l_1,\l_2)\int d^3\hat{x}_+\sqrt{h_+}(\partial^a\phi_{\l_1}^+\partial_a\phi^+_{\l_2} - (1+(\l_1^2+\l_2^2)/2)\phi^+_{\l_1}\phi^+_{\l_2}) \\
        M(\l_1,\l_2) &= \int_0^\infty dt t T_{\lambda_1}(t)T_{\lambda_2}(t).
        \end{split}
\end{equation}
Expanding the integral for $M$, we have that 
\begin{equation}
    \begin{split}
        M &= \int_0^\infty \frac{dt}{t}J_{-i\lambda_1}(mt)J_{-i\lambda_2}(mt) \\
        &= \int_0^\infty\frac{dt}{t}J_{-i\lambda_1}(t)J_{-i\lambda_2}(t).
    \end{split}
\end{equation}
We regulate the integral by multiplying the integrand by $t^\epsilon$; the resulting integral becomes
\begin{equation}
    \begin{split}
        M &= \int_0^\infty\frac{dt}{t^{1-\epsilon}}J_{-i\lambda_1}(t)J_{-i\lambda_2}(t) \\
        &= \frac{\G(1-\epsilon)\G((2-\D_1-\D_2+\epsilon)/2)}{2^{1-\eps}\G((4-\D_1-\D_2-\epsilon)/2)\G((2+\D_1-\D_2-\epsilon)/2)\G((2-\D_1+\D_2-\epsilon)/2)}.
    \end{split}
\end{equation}
This has a pole at $\D_2 = 2-\D_1+\epsilon$; evaluating the $\Delta_2$ integral using the residue theorem therefore gives 
\begin{equation}
    \begin{split}
        S_+ &= -\frac{1}{2}\int [d\Delta]\frac{\sinh\pi\lambda}{\pi\lambda}\int d^3\hat{x}_+\sqrt{h_+}(\partial^a\phi^+_{\lambda}\partial_a\phi^+_{-\lambda} - (1+\lambda^2)\phi^+_\lambda\phi^+_{-\lambda}).
    \end{split}
\end{equation}
Time reversal symmetry then implies that 
\begin{equation}
    \begin{split}
        S_- = -\frac{1}{2}\int [d\Delta]\frac{\sinh\pi\lambda}{\pi\lambda}\int d^3\hat{x}_-\sqrt{h_-}(\partial^a\phi^-_{\lambda}\partial_a\phi^-_{-\lambda} - (1+\lambda^2)\phi^-_\lambda\phi^-_{-\lambda}).
    \end{split}
\end{equation}
\subsection{Lightcone Terms}
Similar to the massless case, the massive action will have terms living on the lightcone arising from discontinuities across $x^2  = 0$. Because the mass term involves no derivatives, it will not contribute to the lightcone action, implying that 
\begin{equation}
    \begin{split}
        S_{L+} &= \int\frac{r^2dr d^2\hat{n}}{2}(\phi^+_{L+}-\phi^0_{L+})\partial_r(\phi^+_{L+}+\phi^0_{L+}) \\
        S_{L-} &= \int\frac{r^2dr d^2\hat{n}}{2}(\phi^-_{L-}-\phi^0_{L-})\partial_r(\phi^-_{L-}+\phi^0_{L-}) 
    \end{split}
\end{equation}
where $\phi^{\pm,0}_{L\pm}$ are defined by taking the appropriate limits from each region of spacetime. Inserting the expansion of the field on the lightcone gives
\begin{equation}
    \begin{split}
        S_{L+} &= -\frac{1}{16\pi}\int [d\Delta]_+\int d^2\hat{n}\frac{1}{\lambda}(2\alpha^+_{-\l}(\pi \l\alpha^0_\l+2\sinh\pi\l\alpha^+_\l) + \pi(2\l\alpha^+_\l - \pi\csch\pi\l\alpha^0_\l)\beta^0_\l) \\
        S_{L-} &=  -\frac{1}{16\pi}\int [d\Delta]_+\int d^2\hat{n}\frac{1}{\lambda}(2\alpha^-_{-\l}(\pi \l\talpha^0_\l+2\sinh\pi\l\alpha^-_\l) + \pi(2\l\alpha^-_\l - \pi\csch\pi\l\talpha^0_\l)\tbeta^0_\l).
    \end{split}
\end{equation}
The full reduced action is therefore 
\begin{empheq}[box=\fbox]{equation}
\begin{split}
    S &= -\frac{1}{2}\int [d\Delta]\frac{\sinh\pi\lambda}{\pi\lambda}\int d^3\hat{x}_+\sqrt{h_+}(\partial^a\phi^+_{\lambda}\partial_a\phi^+_{-\lambda} - (1+\lambda^2)\phi^+_\lambda\phi^+_{-\lambda}) \\
    &-\frac{1}{2}\int [d\Delta]_+\frac{\pi}{4\lambda\sinh\pi\lambda}\int d^3\hat{x}\sqrt{h}[\partial^a\phi^0_\lambda\partial_a\phi^0_\lambda + (1+\lambda^2)\phi^0_\lambda\phi^0_\lambda] \\
    &-\frac{1}{2}\int [d\Delta]\frac{\sinh\pi\lambda}{\pi \lambda}\int d^3\hat{x}_-\sqrt{h_-}(\partial^a\phi^-_{\lambda}\partial_a\phi^-_{-\lambda} - (1+\lambda^2)\phi^-_\lambda\phi^-_{-\lambda}) \\
    & -\frac{1}{16\pi}\int [d\Delta]_+\int d^2\hat{n}\frac{1}{\lambda}(2\alpha^+_{-\l}(\pi \l\alpha^0_\l+2\sinh\pi\l\alpha^+_\l) + \pi(2\l\alpha^+_\l - \pi\csch\pi\l\alpha^0_\l)\beta^0_\l)\\
       &-\frac{1}{16\pi}\int [d\Delta]_+\int d^2\hat{n}\frac{1}{\lambda}(2\alpha^-_{-\l}(\pi \l\talpha^0_\l+2\sinh\pi\l\alpha^-_\l) + \pi(2\l\alpha^-_\l - \pi\csch\pi\l\talpha^0_\l)\tbeta^0_\l).
\end{split}
\end{empheq}
\section{Discussion}
In this paper we have defined a set of coupled actions on 3D AdS and dS spaces that replicate four-dimensional flat physics in Minkowski and Klein space. For massless particles, the asymptotic limits of the reduced field in both Minkowski and Klein space simply encode the values of the 4D field on the lightcone and null infinity, showing that familiar AdS$_3$ extrapolate dictionaries can reproduce the familiar null infinity limits of Carrollian holograms for massless scattering \cite{Ruzziconi:2026bix}.  

This reduction may be useful in several ways for constructing flat space holograms from the AdS$_3$/CFT$_2$ dictionary. 
\begin{itemize}
    \item[1)] When performing quantization, one chooses a set of lower-dimensional slices on which to define a state. If these sets of `constant time slices' can be smoothly deformed into one another, the resulting Hilbert spaces are unitarily equivalent, but if the constant time slices are not smoothly deformable into each other, the resulting Hilbert space may be distinct. This has already been used to generate representations of the Kleinian $S$ matrix where translation invariance emerges from entanglement between two quantum systems \cite{Melton:2024pre}. In (2,2) signature Klein space, the time direction on each unit hyperbolic slice is an angular direction in the 4D spacetime, so quantizing the reduced action in Equation \eqref{eq:kleinact} will generate a distinct Hilbert space that may be more amenable to a holographic description. While time on the hyperbolic slice is periodic, it has recently been shown that certain free theories can be consistently quantized on AdS$_3/\mathbb{Z}$ \cite{Melton:2025ecj}. In fact, the Kleinian reduced action found here has an off-diagonal kinetic term connecting modes of weight $1+i\lambda$ and $1-i\lambda$, which was necessary for defining a nonnegative inner product on the set of single-valued solutions of the wave equation in Klein space \cite{Melton:2025ecj}. 
    \item[2)] Recently, we have understood how to extract distributional celestial MHV amplitudes from linear combinations of celestial leaf amplitudes \cite{Melton:2023bjw, Melton:2024akx}. Beyond the MHV sector, bulk propagation needs to be taken into account, and the lightcone terms in the reduced actions in Equations \eqref{eq:minkmasslessaction} and \eqref{eq:kleinact} tell us how to combine AdS$_3$ and dS$_3$ bulk-to-boundary propagators to replicate bulk-to-bulk propagation between the timelike and spacelike regions of Minkowski and Klein space.            
    \item[3)] Flat space holography has struggled to describe massive particles, which naturally reach $i^\pm$ rather than $\scri^\pm$. While the 3D extrapolate limits for massless fields replicate standard extrapolate dictionaries for massless particles in 4D, the 4D interpretation of the extrapolate dictionary for the reduced massive modes is much less clear. Understanding the precise interpretation of the 3D extrapolate dictionary applied to the reduced massive field therefore promises to help describe massive particles holographically.
\end{itemize}

\acknowledgments

The author would like to thank Simon Heuveline, Lionel Mason, Nia Robles, Romain Ruzziconi, Ahmed  Sheta, Andrew Strominger, Tianli Wang, and Hongji Wei for useful conversations. WM is supported by a Junior Fellowship from the Society of Fellows of Harvard University. Generative AI models were used for copy editing, checking equations, and for clarifying minor technical points. All results and computations have been verified by hand.


\bibliographystyle{JHEP}
\bibliography{biblio.bib}


\end{document}